# Self-supervised Deep Learning for Denoising in Ultrasound Microvascular Imaging

Lijie Huang, Jingyi Yin, Jingke Zhang, U-Wai Lok, Ryan M. DeRuiter, Jieyang Jin, Kate M. Knoll, Kendra E. Petersen, James D. Krier, Xiang-yang Zhu, Gina K. Hesley, Kathryn A. Robinson, Andrew J. Bentall, Thomas D. Atwell, Andrew D. Rule, Lilach O. Lerman, Shigao Chen, and Chengwu Huang

*Abstract*—Ultrasound microvascular imaging (UMI) is often hindered by low signal-to-noise ratio (SNR), especially in contrast-free or deep tissue scenarios, which impairs subsequent vascular quantification and reliable disease diagnosis. To address this challenge, we propose Half-Angle-to-Half-Angle (HA2HA), a self-supervised denoising framework specifically designed for UMI. HA2HA constructs training pairs from complementary angular subsets of beamformed radio-frequency (RF) blood flow data, across which vascular signals remain consistent while noise varies.

HA2HA was trained using *in-vivo* contrast-free pig kidney data and validated across diverse datasets, including contrast-free and contrast-enhanced data from pig kidneys, as well as human liver and kidney. An improvement exceeding 15 dB in both contrast-to-noise ratio (CNR) and SNR was observed, indicating a substantial enhancement in image quality. In addition to power Doppler imaging, denoising directly in the RF domain is also beneficial for other downstream processing such as color Doppler imaging (CDI). CDI results of human liver derived from the HA2HA-denoised signals exhibited improved microvascular flow visualization, with a suppressed noisy background. HA2HA offers a label-free, generalizable, and clinically applicable solution for robust vascular imaging in both contrast-free and contrast-enhanced UMI.

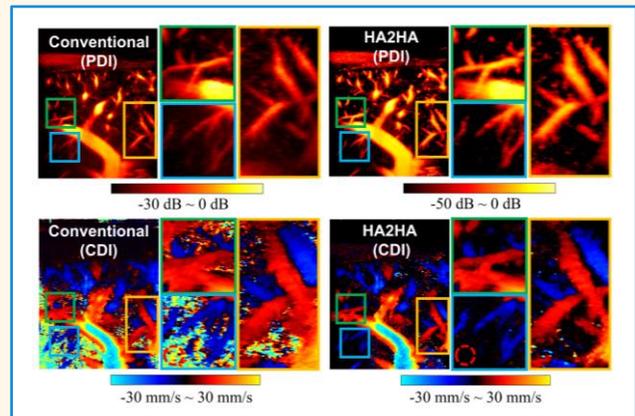

## I. Introduction

ULTRASOUND microvascular imaging (UMI) has emerged as a critical modality for visualizing the microcirculatory system [1, 2], playing a vital role in diagnosing and monitoring various pathological conditions. It can be broadly classified into contrast-free (i.e., without using ultrasound contrast agents) and contrast-enhanced imaging methods. Contrast-free UMI relies on detecting backscattered ultrasound signals from red blood cells, while contrast-enhanced UMI employs intravenously injected microbubble (MB) contrast agents, which exhibit strong acoustic scattering properties. These contrast agents can be further localized and tracked to achieve super-resolution vascular imaging (i.e., ultrasound localization microscopy, ULM) [3-5]. However, despite these advancements, image quality in UMI remains a major challenge, particularly in deep regions, due to the inherently low signal-to-noise ratio (SNR) of unfocused plane wave imaging. The low SNR, combined with weak blood flow signals in small vessels, makes it difficult to effectively distinguish blood flow signals from background noise. Consequently, the performance of blood flow imaging is compromised, particularly for flow speed estimation and microbubble localization, ultimately limiting reliable

Research reported in this publication was supported by the National Institute of Diabetes and Digestive and Kidney Diseases under Award Number R01DK129205. The content is solely the responsibility of the authors and does not necessarily represent the official views of the National Institutes of Health.

(*Corresponding authors: Shigao Chen and Chengwu Huang*)

L. Huang, J. Yin, J. Zhang, U.-W. Lok, R.M. DeRuiter, J. Jin, K.M. Knoll, K. E. Petersen, G.K. Hesley, K. A. Robinson, T. D. Atwell, *C. Huang, and *S. Chen are with the Department of Radiology, Mayo Clinic College of Medicine and Science, Rochester, MN 55905 USA (e-mail: Chen.Shigao@mayo.edu; Huang.Chengwu@mayo.edu).

J. Jin was with the Department of Radiology, Mayo Clinic College of Medicine and Science, Rochester, MN 55905 USA. She is now with the Department of Ultrasound, the Third Affiliated Hospital of Sun Yat-Sen University, 600 Tianhe Road, Guangzhou, P.R. China.

J. D. Krier, X.Y. Zhu, A.J. Bentall, A.D. Rule, L. O. Lerman are with the Division of Nephrology and Hypertension, Mayo Clinic, Rochester, MN 55905 USA.

> **Highlights**
> - A self-supervised framework (HA2HA) leverages angular subset pairs of blood flow RF data to construct training inputs for denoising ultrasound microvascular imaging (UMI) without clean labels.
> - Direct RF-domain denoising improves both power Doppler and color Doppler imaging, achieving >15 dB gain in CNR and SNR across diverse datasets.
> - HA2HA provides a generalizable, label-free solution for robust ultrasound microvascular flow imaging, with strong potential for clinical translation.

microvascular quantification and clinical applications.

To address this limitation, several signal processing techniques have been developed, such as adaptive beamforming [6-11], compounding [10, 12-14], advanced clutter filtering [2, 15-19] and post-processing methods [20-24]. However, most adaptive beamformers involve high computational complexity and require access to raw channel data. In addition to computational cost, advanced clutter filtering methods and post-processing techniques usually require extensive and application-specific parameter tuning, limiting their generalizability.

Supervised deep learning has been used in ultrasound imaging denoising [5], which utilizes labeled datasets to enable efficient and precise mapping between noisy and clean images. However, obtaining clean reference images *in-vivo* for ultrasound imaging is often impractical due to the presence of inherent noise, complex tissue interactions, and variability across different imaging conditions.

In contrast, self-supervised denoising approaches offer a promising alternative by learning directly from noisy data without requiring clean labels. These methods exploit the statistical structure of noise and redundancy within the data itself. This approach is particularly advantageous in medical imaging scenarios where high-quality ground truth data is difficult or expensive to obtain.

Self-supervised denoising methods are typically categorized into single-input [25-27] and paired-input [28] approaches. The single-input strategy requires only a single noisy image and relies on spatial masking or data redundancy, while the paired-input approach trains on independently acquired noisy observations of the same underlying signal under the assumption that the noise in both observations is independent and zero-mean.

Among them, Noise2Noise (N2N) is a leading paired-input strategy particularly well-suited to medical imaging, where repeated acquisitions can be leveraged to form noisy input pairs. This enables effective training without the need for clean ground truth, which is rarely available in clinical practice. N2N-based frameworks have demonstrated success across a variety of imaging modalities, including MRI [29], CT [30, 31], optical imaging [32, 33], photoacoustic imaging [34], and PET [35].

In ultrasound imaging, N2N-based methods have primarily focused on suppressing speckle noise in B-mode images, where paired noisy inputs are constructed either by adding noise [36] to simulation data or by selecting temporally adjacent frames [37]. Goudarzi and Rivaz [38] applied the N2N principle directly to consecutive frames of ultrasound radio-frequency (RF) data, demonstrating improved imaging quality in deep regions.

Despite these advances, the application of N2N to UMI remains largely unexplored. UMI often relies on angularly compounded plane wave acquisitions, in which each transmit angle inherently introduces a distinct noise realization within the same acquisition frame. This angular diversity naturally provides multiple observations of the same underlying signal with independent noise, making it well suited for constructing paired noisy inputs without relying on simulation data or temporally paired frames, potentially improving the generalization and mitigating issues like frame misalignment caused by tissue motion.

In this study, we propose Half-Angle-to-Half-Angle (HA2HA), a self-supervised denoising framework specifically designed for UMI. Inspired by the N2N paradigm, HA2HA constructs training pairs by splitting the steered angles of beamformed RF blood flow data into two complementary half-angle subsets (angular subsets), where vascular signals remain consistent while noise varies. Then, each subset was compounded and clutter filtered separately prior to training, allowing the network to focus on vascular signal denoising while minimizing tissue interference. Unlike image-domain or envelope-based methods, HA2HA operates directly on RF signals, which is beneficial for downstream tasks such as color Doppler imaging (CDI) that leverages the phase information of the denoised RF signal.

HA2HA was trained on *in-vivo* contrast-free pig kidney data and validated across diverse datasets, including data from contrast-free and contrast-enhanced pig kidneys, as well as data from human liver and kidney. This variation in imaging conditions, anatomical regions, and species enabled a comprehensive evaluation of the method's robustness and adaptability across a wide range of clinical and experimental scenarios. To evaluate the effectiveness of the proposed method, quantitative comparisons were conducted against the conventional method, angular processing (AP), and spatiotemporal non-local means filtering (ST-NLM). The performance was assessed using contrast-to-noise ratio (CNR), SNR, and background noise power (BNP). An improvement exceeding 15 dB in both CNR and SNR was observed, offering improved visualization of microvascular structures. In addition to power Doppler imaging, CDI was also performed based on the HA2HA denoised blood flow RF signals. The resulting CDI map, computed from the denoised human liver data, demonstrated improved flow speed visualization with suppression of noisy background.

In summary, we propose HA2HA, a label-free and self-supervised denoising framework for blood flow RF signals in UMI. HA2HA leverages angular-domain redundancy to construct training pairs without the need for clean labels. Unlike most existing approaches that operate in the image domain [37], HA2HA performs denoising directly on RF signals, which facilitates improved downstream processing such as CDI that are sensitive to noise in RF signal. The method also shows strong generalizability across contrast-free and contrast-enhanced imaging settings and various anatomical regions.

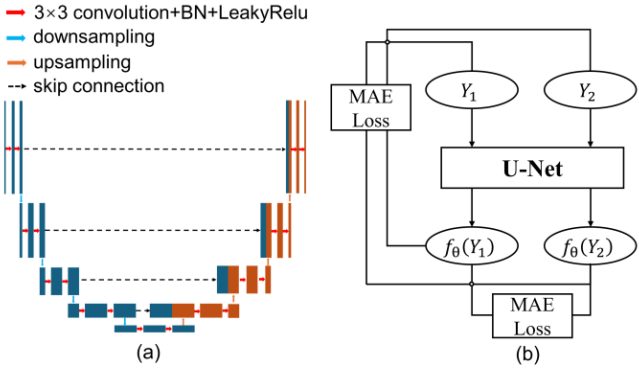

Fig. 1. (a) Network architecture for the proposed HA2HA method. (b) HA2HA self-supervised dual-path framework.

## II. METHODS

### A. Network Architecture

As illustrated in Fig. 1a, we adopt a 2D U-Net architecture as the backbone of our self-supervised denoising framework. U-Net is widely recognized for its effectiveness in biomedical imaging due to its ability to capture both global contextual features and fine spatial details. Our implementation follows a conventional encoder–decoder design with four resolution levels.

Each encoder stage comprises two 3×3 convolutional layers, with each convolution followed by batch normalization and a LeakyReLU activation function. Max pooling is used to progressively reduce spatial resolution and expand the receptive field. In the decoder, bilinear interpolation is used for up-sampling and skip connections from the encoder are incorporated to preserve high-frequency features and stabilize training.

The network is trained using pairs of noisy observations acquired from the same anatomical region. These pairs retain the same underlying vascular structure while exhibiting uncorrelated noise, enabling the network to learn clean signal representations without access to ground-truth labels.

### B. HA2HA Self-supervised Framework

The proposed framework builds on the N2N principle, which enables the training of a neural network for image denoising using only pairs of noisy observations, without requiring clean references. The key insight is that the network can learn to recover the underlying clean signal by mapping one noisy observation to another independent noisy observation of the same scene. Let $Y_1 = X + N_1$ and $Y_2 = X + N_2$, where $X$ denotes the unknown clean signal, $N_1$, $N_2$ are independent realizations of zero-mean noise. In the context of HA2HA, the term "noise" primarily refers to stochastic disturbances (e.g., thermal or electronic noise), but also encompasses structured interference such as angle-dependent sidelobes and clutter variations arising from complementary half-angle compounding. For simplicity, we refer to all such components collectively as "noise" throughout the paper.

Given that $Y_1$ and $Y_2$ share the same underlying signal $X$ but differ in noise, the network $f_\theta$ is trained to approximate the underlying clean signal $X$ by minimizing the expected loss:

$$\mathcal{L}(\theta) = \mathbb{E}[d(f_\theta(Y_1), Y_2)] \quad (1)$$

where $d(\cdot)$ is a distance metric, commonly chosen as the mean absolute error (MAE) or the mean squared error (MSE), and E denotes the expectation over all training pairs. In this study, the MAE-based objective is used:

$$\mathcal{L}(\theta) = \mathbb{E}[|f_\theta(Y_1) - Y_2|] \quad (2)$$

The fundamental assumption underlying HA2HA is that the noise components $N_1$ and $N_2$ are independent and zero-mean, an assumption that is generally valid for noise in ultrasound RF signals:

$$\mathbb{E}[Y_2|Y_1] = X + \mathbb{E}[N_2|Y_1] = X \quad (3)$$

As a result, minimizing the loss with respect to $Y_2$ implicitly guides the network output $f_\theta(Y_1)$ toward the clean signal $X$, despite the absence of explicit supervision. This theoretical foundation supports the feasibility of training denoising models in a fully self-supervised manner. The specific construction of the proposed HA2HA training data is described in detail in Section 2.6.

### C. Loss Function and Training Objectives

As illustrated in Fig. 1b, the HA2HA loss consists of three components: (1) a forward loss between the network output and its noisy target, (2) a reverse loss using the second input as the network input and the first as the target, and (3) a consistency loss enforcing similarity between the two outputs. This composite structure encourages the network to extract the common underlying signal while suppressing independent noise in each input. The total HA2HA loss is given by:

$$\mathcal{L}_{N2N} = \frac{|f_\theta(Y_1) - Y_2|_1 + |f_\theta(Y_2) - Y_1|_1 + \lambda_c \cdot |f_\theta(Y_1) - f_\theta(Y_2)|_1}{2 + \lambda_c} \quad (4)$$

Here, $\lambda_c \in [0,1]$ is a weighting coefficient that enforces structural consistency between the two denoised outputs. In this study, $\lambda_c$ is set to 0.5 based on empirical tuning (see Supplementary Fig. S1). We adopt the MAE as the primary training objective due to its robustness to outliers and its ability to preserve sparse vascular structures while suppressing noise. In addition to the denoising loss, we apply $L_1$ regularization to the network parameters to mitigate overfitting. The final loss is defined as:

$$\mathcal{L}_{total} = \mathcal{L}_{N2N} + \lambda_1 |\theta|_1 \quad (5)$$

where $\theta$ denotes the network parameters, and $\lambda_1$ control the strength of the regularization terms. In our experiments, $\lambda_1$ is set to $1 \times 10^{-5}$, selected based on empirical performance (see

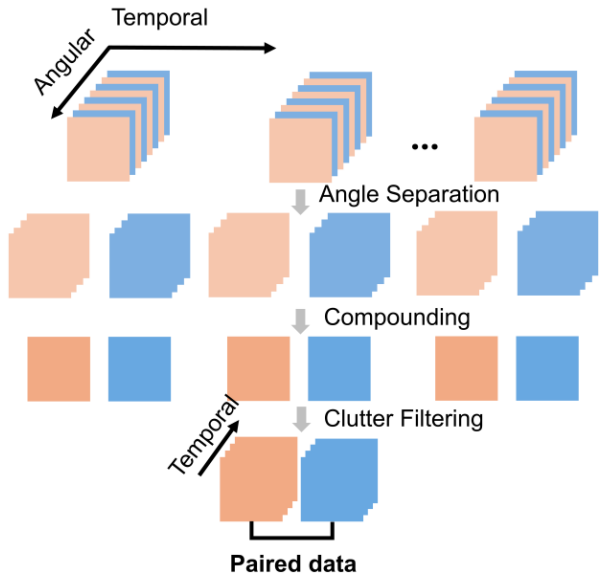

Fig. 2. Construction of paired RF training inputs.

Supplementary Fig. S2). This combination allows the model to learn robust denoising mappings from noisy data while maintaining compactness and generalizability.

### D. Implementation Details

The model is implemented in PyTorch. It is trained with a batch size of 256 on 7,200 paired blocks of size 128×128. Training is performed using the AdamW optimizer ($\beta_1 = 0.5$, $\beta_2 = 0.999$) [39], with an initial learning rate of $1 \times 10^{-4}$. A reduce-on-plateau learning rate schedule is employed, decreasing the learning rate by a factor of 0.5 when the training loss plateaued for 10 epochs.

The model converges in approximately 2.5 hours on an NVIDIA Quadro RTX 5000 GPU. During inference, each 3D blood flow RF dataset (1433×1152×300; axial × lateral × temporal), with a pixel resolution of $\frac{\lambda}{8} \times \frac{\lambda}{8}$ in both axial and lateral directions, is processed in a frame-wise manner, where each 2D RF frame is independently denoised. The entire volume is processed in approximately 67 seconds.

### E. Data Acquisition

All the *in-vivo* data was acquired using Verasonics Vantage 256 system (Verasonics Inc., Kirkland, WA, USA) equipped with a GE9L-D linear array probe (GE Healthcare, USA). The imaging parameters of these data will be introduced in detail.

#### 1) Pig kidney, human liver and kidney study

Approved by the Institutional Animal Care and Use Committee (IACUC) of Mayo Clinic, the pig kidney data was acquired from several pigs, and approved by the Institutional Review Board (IRB) of Mayo Clinic, the human liver data was acquired from the right liver of a healthy volunteer, and the human kidney data were acquired from a patient with chronic kidney disease (CKD). Informed consent was obtained from the patient. The transmit center frequency was 5.208 MHz, and the sampling frequency was 20.832 MHz. The pulse repetition frequency (PRF) was 5 kHz, and ten plane waves with steering angles of -9° to 9° at an interval of 2° were transmitted. And the effective frame rate was 500 Hz. A total of 300 frames of raw channel RF data were acquired for each dataset.

#### 2) Pig kidney acquired under different SNRs

To obtain ultrasound blood flow data under varying SNR conditions, the pig kidney was scanned using different acoustic output by modulating the duty cycle (DC) of transmission. Four DC values—0.1, 0.2, 0.4, and 0.8—were used to mimic different levels of transmit energy. Six steering angles, ranging from −7.5° to 7.5° with a step size of 3°, were applied for plane wave compounding. The pulse repetition frequency (PRF) was 6667 Hz, and the effective frame rate was 250 Hz.

To ensure consistent imaging across different DC levels within each frame, the acquisitions were organized as follows: for each frame, the six steering angles were first transmitted under DC = 0.1, followed by six angles under DC = 0.2, then 0.4, and finally 0.8. This sequential DC-based acquisition ensured that all transmit energy levels shared the same imaging section and motion status.

### F. Training and Inference Data Processing

As shown in Fig. 2, steered plane wave angles were divided into odd and even groups. For each group, coherent compounding was performed independently, followed by singular value decomposition (SVD)-based clutter filtering, resulting in two complementary blood flow RF subsets.

The training dataset consisted of 90 frames of *in-vivo* contrast-free pig kidney RF data acquired from three pigs. For each pig, 30 frames were uniformly sampled from a 300-frame ensemble (one every 10 frames) from a representative imaging plane, to enhance vascular variability and improve training diversity. Each RF frame initially had a pixel size of $\frac{\lambda}{8} \times \frac{\lambda}{2}$, where $\lambda$ is the acoustic wavelength. After lateral interpolation to achieve isotropic resolution, the pixel size became $\frac{\lambda}{8} \times \frac{\lambda}{8}$. The interpolated frame was then divided into non-overlapping $128 \times 128$ patches with a stride of 128, yielding 80 paired patches per frame. Basic spatial augmentations, including random flipping and rotation, were applied to each patch pair, yielding a total of 7,200 paired training samples.

To assess model generalizability, additional full-angle compounded datasets were used, including *in-vivo* contrast-free and contrast-enhanced pig kidney, as well as human liver and kidney RF data. During inference, the model was applied directly to full-size clutter-filtered RF frames without patch-wise tiling, preserving spatial continuity and avoiding boundary artifacts. Single-frame B-mode and power Doppler images were generated from clutter-filtered RF data before and after HA2HA denoising for comparison.

### G. Compared Methods

The conventional method reconstructs power Doppler images from full-angle coherently compounded data after SVD-based clutter filtering. The AP method exploits the incoherent nature of noise across different steering angles to achieve effective noise suppression [14]. Specifically, the steered plane wave angles are divided into odd and even groups, similar to the training data construction in the proposed HA2HA framework. Each group is independently compounded and clutter-filtered using SVD. The resulting RF signals are then converted into complex-valued signals via Hilbert transform and conjugately multiplied to enhance flow components while attenuating angle-incoherent noise. The ST-NLM filtering method exploits the differing trajectories of continuous blood flow and temporally uncorrelated noise [20] on the spatiotemporal (axial-temporal) domain. It was applied to the full-angle coherently compounded and SVD-filtered blood flow RF data. The similarity window size was set to 11 × 11 (384 μm × 22 ms), and the search window was defined as twice the similarity window in both spatial and temporal dimensions.

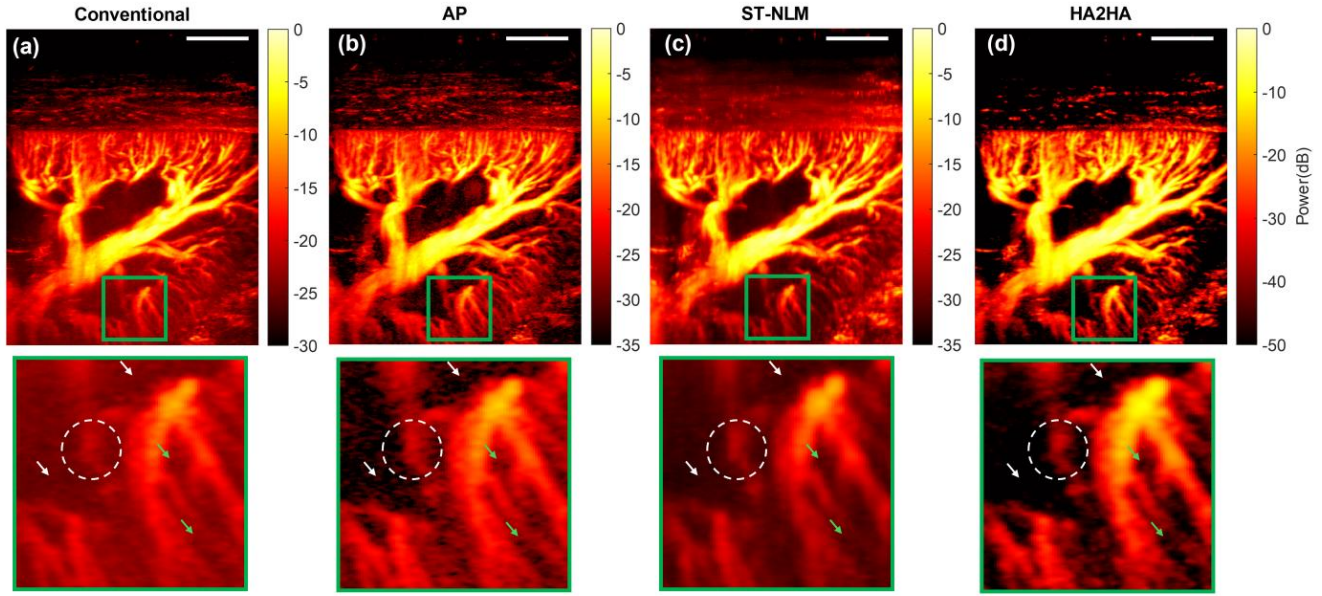

Fig. 3. Power Doppler images of *in-vivo* contrast-free pig kidney using (a) the conventional method, (b) AP, (c) ST-NLM and (d) the proposed HA2HA method. The white horizontal line represents 1 cm.

### H. Evaluation Metrics

Quantitative evaluation of the proposed HA2HA framework was conducted on final power Doppler images using three metrics: CNR, SNR and BNP. Among these, CNR quantifies contrast between blood flow and background, while SNR and BNP assess noise suppression performance. The CNR is calculated as:

$$\text{CNR} = 10 \times \log_{10}\left(\frac{\bar{S}_{blood} - \bar{S}_{background}}{\sigma_{noise}}\right) [dB] \quad (6)$$

where $\bar{S}_{blood}$ and $\bar{S}_{background}$ denote the mean intensities of blood flow and background regions, respectively. $\sigma_{noise}$ is the standard deviation of a region containing only noise. The SNR is computed as:

$$\text{SNR} = 10 \times \log_{10}\left(\frac{\bar{S}_{blood}}{\sigma_{noise}}\right) [dB] \quad (7)$$

The BNP is defined as:

$$\text{BNP} = 10 \times \log_{10}(\bar{S}_{noise}) [dB] \quad (8)$$

where $\bar{S}_{noise}$ is the mean intensity of the same noise-only region used for computing $\sigma_{noise}$. The regions-of-interest (ROIs) definitions for blood flow, background, and noise regions used in the above quantitative metrics are provided in Supplementary Fig. S3 for reference.

## III. RESULTS

To assess generalizability, the HA2HA network was trained only once on paired *in-vivo* contrast-free pig kidney blood flow RF data. It was then evaluated without any fine-tuning on multiple datasets, including *in-vivo* contrast-enhanced pig kidney, contrast-free human liver, and human kidney with CKD. Additionally, to test the proposed method under varying SNR conditions, contrast-free pig kidney data acquired using different DCs were also employed.

### A. In-vivo contrast-free pig kidney

Fig. 3 shows the power Doppler images of a contrast-free pig kidney obtained using the conventional method, AP, ST-NLM and the proposed HA2HA method. The displayed dynamic range of each method was individually optimized to balance microvascular visibility background noise. The zoomed-in views of the green boxes illustrate that HA2HA achieves lower background noise between microvascular branches than compared methods, as indicated by the green arrows. The white dotted circle highlights the improved delineation of fine vessels enabled by the proposed method. In the pure noise region, marked by the white arrows, HA2HA consistently achieves the lowest noise level. Quantitative results in Table 1 further confirm the superior performance of HA2HA, demonstrating the best CNR, SNR and BNP values among all methods.

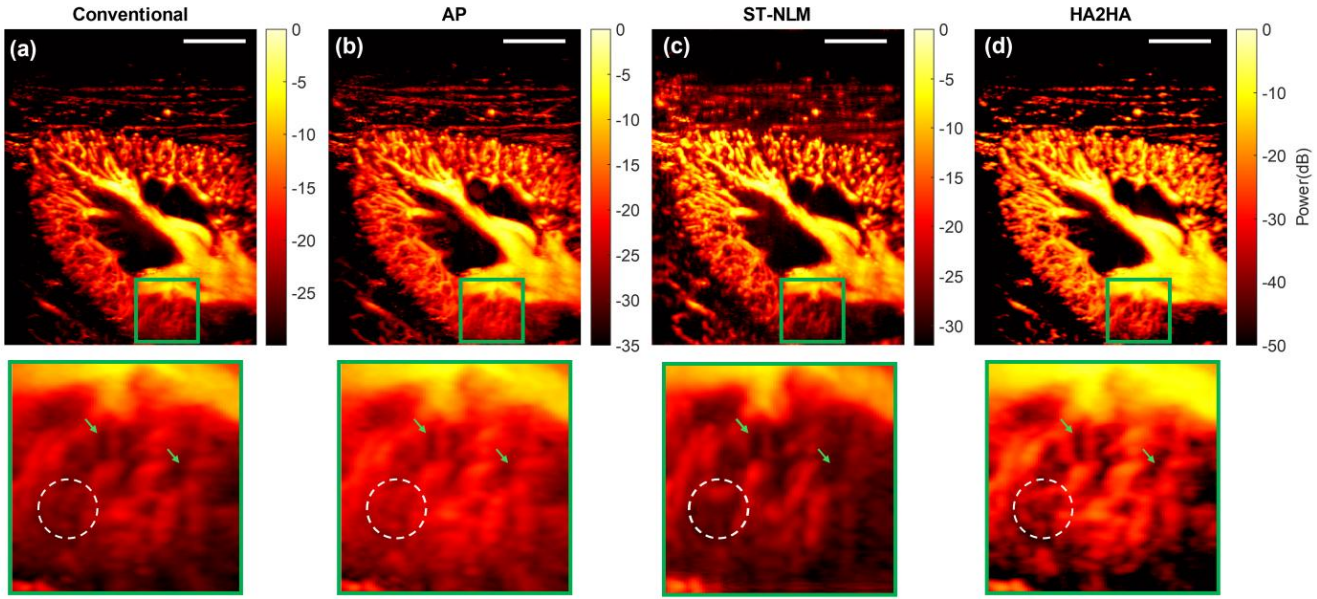

Fig. 4. Power Doppler images of *in-vivo* contrast-enhanced pig kidney using (a) the conventional method, (b) AP, (c) ST-NLM and (d) the proposed HA2HA method. The white horizontal line represents 1 cm.

TABLE I
QUANTITATIVE RESULTS ON CONTRAST-FREE AND CONTRAST-ENHANCED PIG KIDNEY DATA USING THE PROPOSED HA2HA METHOD AND THE COMPARED METHODS

| Indexes | Pig Kidney | | | | | | | |
|---|---|---|---|---|---|---|---|---|
| | Contrast-free | | | | Contrast-enhanced | | | |
| | Conventional | AP | ST-NLM | HA2HA | Conventional | AP | ST-NLM | HA2HA |
| CNR [dB] | 6.38 | 12.84 | 12.89 | **27.82** | 6.15 | 12.49 | 5.92 | **17.57** |
| SNR [dB] | 7.28 | 13.06 | 12.90 | **27.82** | 7.10 | 12.73 | 6.06 | **17.64** |
| BNP [dB] | -12.51 | -16.31 | -15.78 | **-24.76** | -14.64 | -18.29 | -15.66 | **-25.94** |

Compared to the single frame of blood flow image obtained by the conventional method, the HA2HA denoised blood flow image exhibits enhanced flow visibility and substantial background noise suppression, especially in deep regions (see Supplementary Video 1).

### B. In-vivo contrast-enhanced pig kidney

Fig. 4 shows the power Doppler images of a contrast-enhanced pig kidney obtained from different methods. As shown by the white dotted circles, vessels delineated by MBs are clearly visualized and easily differentiated from the background when using the HA2HA method, whereas they are difficult to distinguish from surrounding noise with the compared methods. Quantitative results in Table 1 further confirm the superior performance of HA2HA across all evaluated metrics.

Importantly, the HA2HA model was trained solely on contrast-free data, while it is still effective for contrast-enhanced data, demonstrating its strong generalization capability across different contrast conditions. The single frame MB image is shown in Supplementary Video 2, where MBs reconstructed by HA2HA are easier to differentiate from background noise.

### C. In-vivo contrast-free human liver

Fig. 5 shows the power Doppler images of a contrast-free human liver, used exclusively for testing. As highlighted by the yellow arrows in the zoomed-in views, HA2HA enhances the visualization of small vessels with weak intensity that are difficult to resolve in the conventional and ST-NLM images. The AP method improves the visibility of small vessels to some extent but is limited by relatively high background noise. In contrast, HA2HA achieves superior vascular visualization with the lowest background noise, which is also quantitatively validated in Table 2. The effectiveness of HA2HA on the human liver data further demonstrates its strong generalization capability across different anatomical regions, despite being trained solely on contrast-free pig kidney data. The single frame

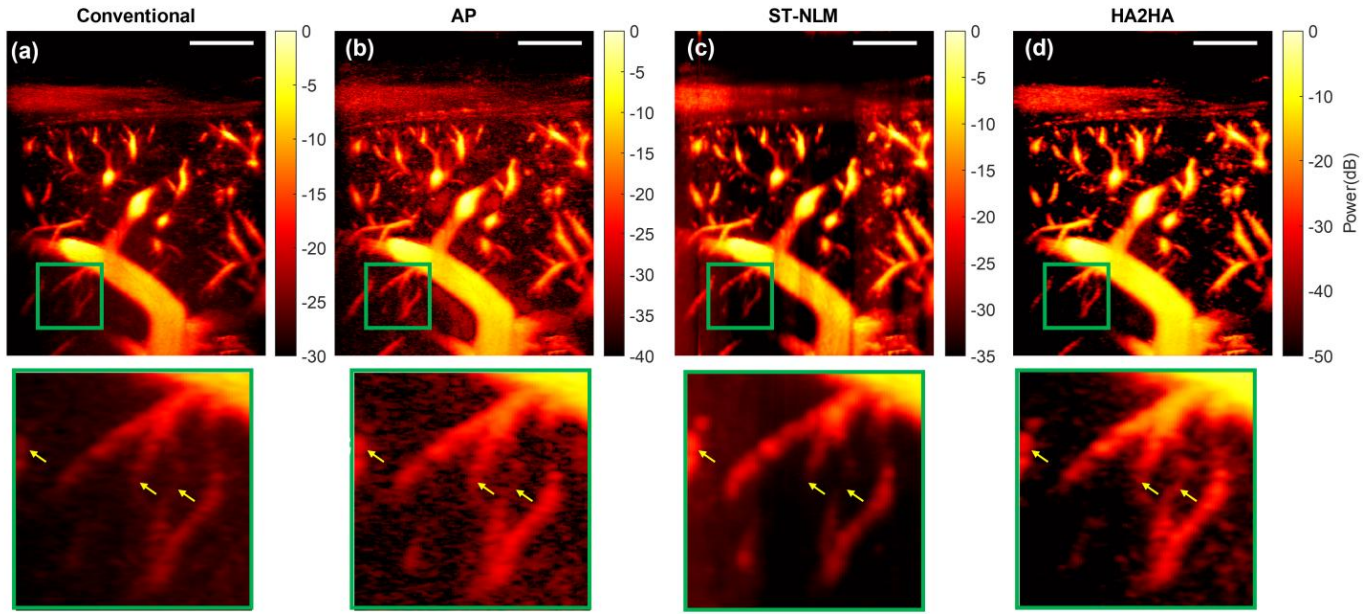

Fig. 5. Power Doppler images of *in-vivo* contrast-free human liver using (a) the conventional method, (b) AP, (c) ST-NLM and (d) the proposed HA2HA method. The white horizontal line represents 1 cm.

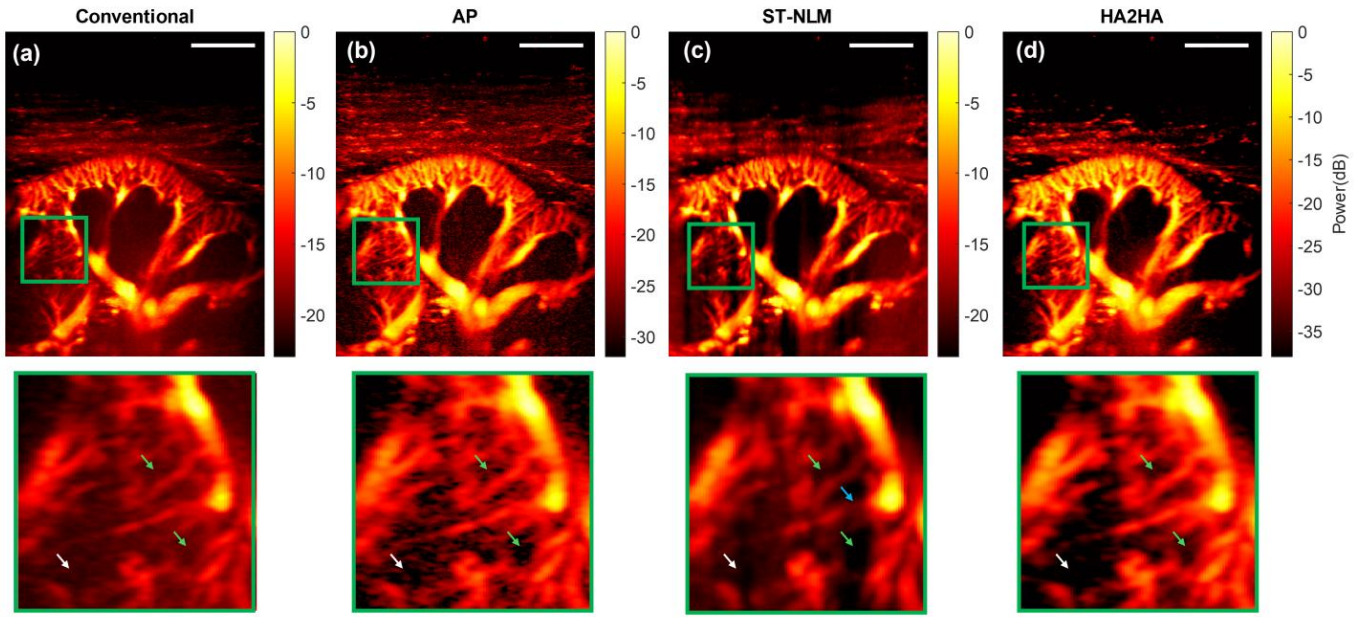

Fig. 6 Power Doppler images of *in-vivo* contrast-free human kidney with CKD using (a) the conventional method, (b) AP, (c) ST-NLM and (d) the proposed HA2HA method. The white horizontal line represents 1 cm.

blood flow images using HA2HA and the conventional method is shown in Supplementary Video 3.

### D. In-vivo contrast-free human kidney

Fig. 6 presents the power Doppler images of a human kidney from a CKD patient. As indicated by the green arrows, HA2HA reduces background noise between adjacent vessels. The white arrows highlight that HA2HA effectively suppresses noise in pure background regions. Notably, as indicated by the blue arrow, ST-NLM shows limited ability to preserve laterally flowing vessels due to its denoising operation being applied in the axial-temporal domain. The quantitative results (Table 2) and the single-frame blood flow image (Supplementary Video 4) both confirm that HA2HA consistently achieves the best performance.

### E. Investigation of DC conditions

The performance of the proposed HA2HA method was further evaluated under systematically varied SNR levels. Contrast-free pig kidney data were acquired using different transmit DCs (DC = 0.8, 0.4, 0.2, and 0.1). Rows 1 to 4 in Fig. 7 show power Doppler images reconstructed using the conventional method, AP, ST-NLM, and HA2HA, respectively, at each DC level. The dynamic range for each image was individually optimized to balance noise suppression and microvascular visibility. As DC decreases, all methods exhibit progressively increased background noise, particularly at DC = 0.2 and 0.1, where small vessels with relatively low intensity become difficult to distinguish from the elevated noise background.

TABLE II
QUANTITATIVE RESULTS ON CONTRAST-FREE HUMAN LIVER AND HUMAN KIDNEY DATA USING THE PROPOSED HA2HA METHOD AND THE COMPARED METHODS

| Indexes | Contrast-free Human Liver | | | | Contrast-free Human Kidney | | | |
|---|---|---|---|---|---|---|---|---|
| | Conventional | AP | ST-NLM | HA2HA | Conventional | AP | ST-NLM | HA2HA |
| CNR [dB] | 5.26 | 12.70 | 6.14 | **25.18** | 4.18 | 12.37 | 8.79 | **19.37** |
| SNR [dB] | 6.39 | 12.93 | 6.26 | **25.20** | 5.58 | 12.62 | 8.83 | **19.42** |
| BNP [dB] | -15.00 | -19.73 | -15.37 | **-27.37** | -10.67 | -15.27 | -11.76 | **-19.09** |

Fig. 8 shows the quantitative comparison of different methods across various DC values. As the DC increases, all methods exhibit improved performance. HA2HA achieves the highest CNR, the highest SNR and the lowest BNP when DC ≥ 0.2, while AP outperforms HA2HA in the extremely low-SNR case (DC = 0.1).

### F. Color Doppler imaging

Fig. 9 illustrates the power Doppler and corresponding CDI results from a contrast-free human liver dataset. Fig. 9a presents the power Doppler image reconstructed using HA2HA, while Fig. 9b–d show the CDI images derived from Doppler phase shifts calculated using the conventional method, ST-NLM, and HA2HA, respectively. CDI was not computed for the AP method due to the angular multiplication-based combination process in AP method is uniquely design for power Doppler estimation only.

As highlighted by the red dashed circles in the zoomed-in views, the CDI image generated from HA2HA-denoised RF data exhibits a homogeneous background with effectively suppressed noise. In contrast, the CDI maps computed from the conventional and ST-NLM methods (Fig. 9b and 9c) display substantial background color noise, with erroneously high velocity values appearing in regions devoid of vascular structures. This suggests that background noise in the RF data can introduce spurious flow velocity estimates in CDI. Moreover, the HA2HA-based CDI image demonstrates improved visual contrast and reduced background clutter compared to the other methods as indicated by white arrows, underscoring the utility of HA2HA in enhancing Doppler image quality and vessel visualization under contrast-free conditions.

Fig. 10 presents the distribution profiles of intensity and velocity along two selected lines indicated in Fig. 9a — green line (Vessel 1) and blue line (Vessel 2). Fig. 10a and c show the power Doppler intensity profiles, while Fig. 10b and d illustrate the corresponding color Doppler velocity profiles.

As shown in Fig. 10a and 10c, the proposed HA2HA method significantly suppresses background noise, thereby enhancing the contrast between blood vessels and surrounding background. This enables clearer vessel boundary definition and facilitates the separation of vascular signals from background noise.

Furthermore, the flow speed profiles in Fig. 10b and 10d demonstrate that HA2HA effectively suppresses spurious velocity artifacts in the background regions, as indicated by the red arrows. In addition, a parabolic flow velocity profile, indicated by the blue arrows, is clearly observed in the HA2HA results. These findings highlight the ability of HA2HA to reduce noise in contrast-free CDI while preserving physiologically meaningful flow signals.

## IV. DISCUSSION

This study presents HA2HA, a self-supervised deep learning framework designed for denoising ultrasound blood flow RF signals. HA2HA learns to recover clean vascular signals without requiring ground-truth data. Paired inputs are constructed from RF data acquired at different plane wave steering angles. Given the assumption that the zero-mean and uncorrelated nature of noise across steering angles, these pairs share similar vascular information while differing in noise, enabling the network to effectively extract the underlying clean signal.

The proposed method offers several key advantages that contribute to its strong performance and practical utility. First, by operating directly on RF signals instead of envelope or image-domain data, the network is particularly suitable for providing denoised RF signals for downstream applications such as CDI that leverage the inherent phase information of ultrasound signals. This leads to improved flow speed mapping, as shown in the CDI results [Fig. 9], and also supports the HA2HA self-supervised learning framework by leveraging the stochastic and approximately zero-mean nature of noise in RF signals, which is essential for effective training. Second, the method constructs training pairs from complementary angular subsets of beamformed RF data, where vascular signals remain consistent, but noise varies between paired data. This enables effective HA2HA training without clean targets. Clutter filtering is applied to both inputs to suppress tissue components and enhance the focus on blood flow structures. Third, the model exhibits strong generalizability across imaging conditions. Despite being trained only on contrast-free pig kidney data, it performs well on unseen datasets, including contrast-enhanced pig kidneys and contrast-free human liver and kidney scans, without requiring fine-tuning.

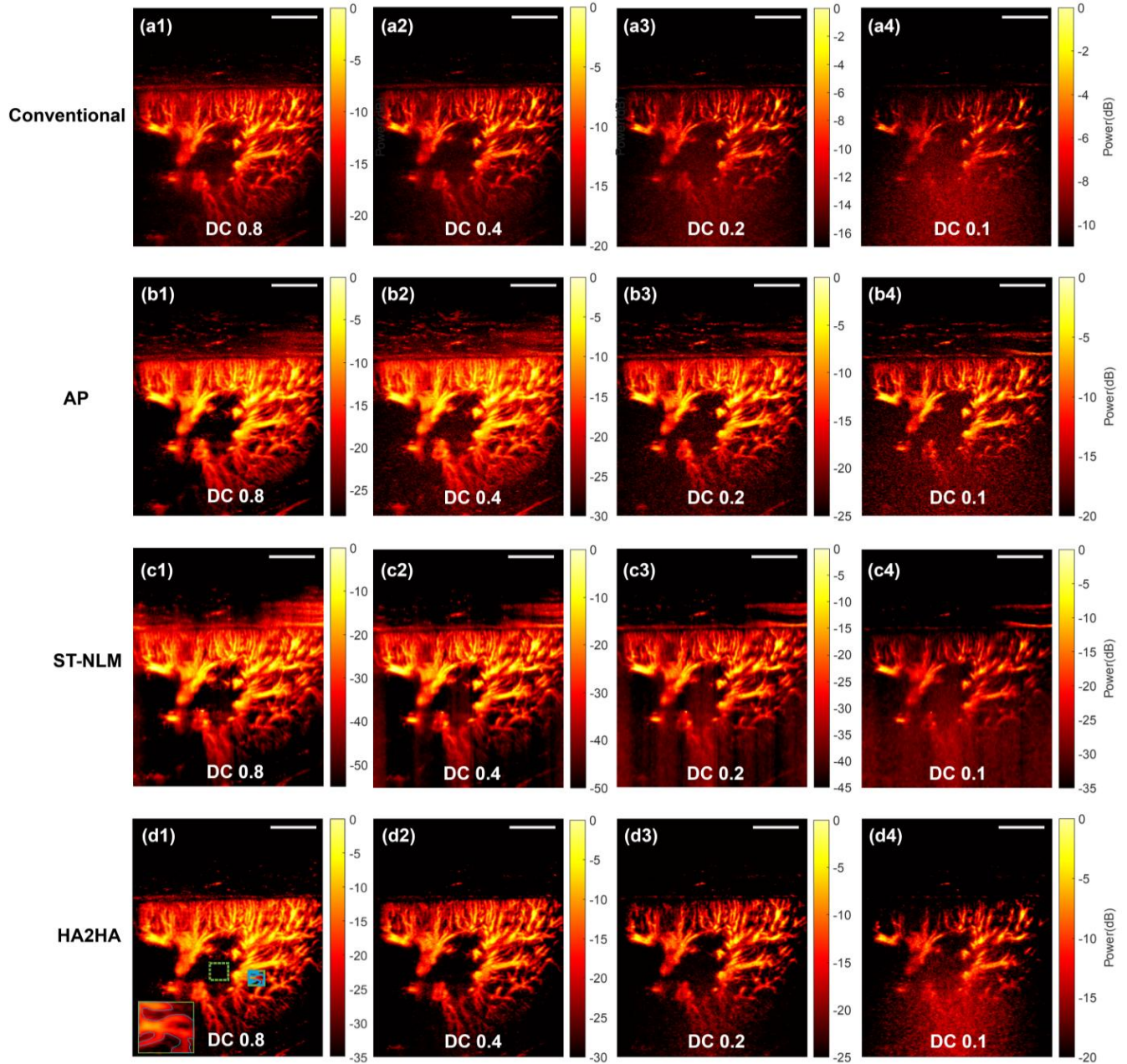

Fig. 7. Power Doppler images of *in-vivo* contrast-free pig kidney obtained by (a1-4) the conventional method, (b1-4) AP, (c1-4) ST-NLM and (d1-4) the proposed HA2HA method under different DC (0.8, 0.4, 0.2, 0.1). The white horizontal line represents 1 cm. In (d1), the ROIs used for quantitative evaluation are defined: the area enclosed by the blue contour represents the manually selected blood flow region; the area outside the blue contour but within the solid green box is defined as the background; and the dashed green box denotes a noise-only region.

The method enables high-quality microvascular imaging without contrast agents. This is especially beneficial for patients with contraindications to contrast-enhanced ultrasound. The improved SNR and CNR facilitate better separation of blood flow signals from background noise, thereby facilitating the microvascular quantification based on power Doppler imaging.

We first evaluated HA2HA by varying the SNR levels of the test data and found that the method remained effective even under low-SNR (e.g., <5 dB), such as the DC = 0.1 case shown in Fig. 8. To assess the impact of training data quality, we trained two models using different types of input pairs: one with high-SNR HA2HA pairs, and the other with low-SNR Single-Angle-to-Single-Angle (SA2SA) pairs constructed from two single-angle acquisitions. Both models were tested on the same high-quality full-angle data. As shown in Supplementary Fig. S4, the model trained on low-quality pairs failed to achieve comparable performance, highlighting the importance of using high-quality data for training.

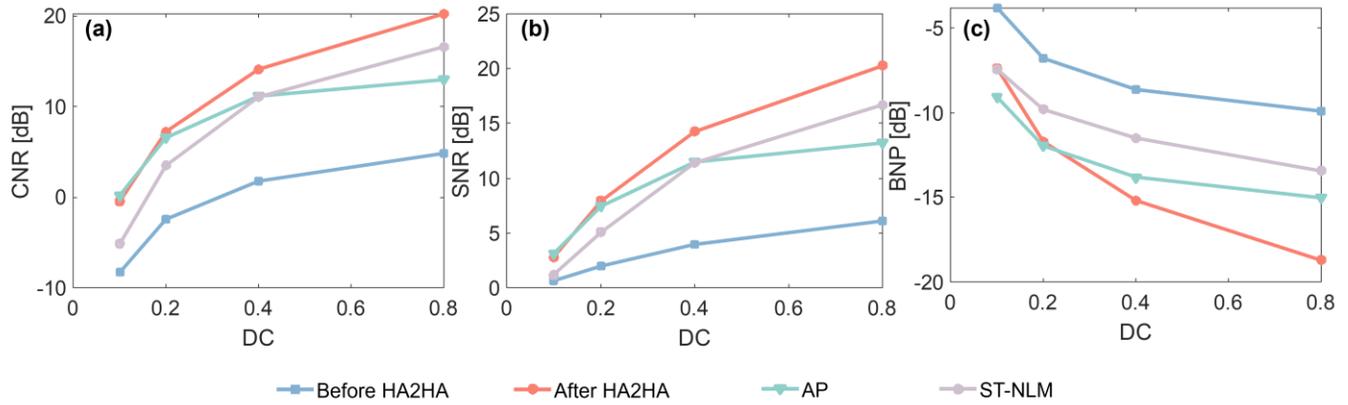

Fig. 8.  (a) CNR, (b) SNR and (c) BNP trends with the change of DC.

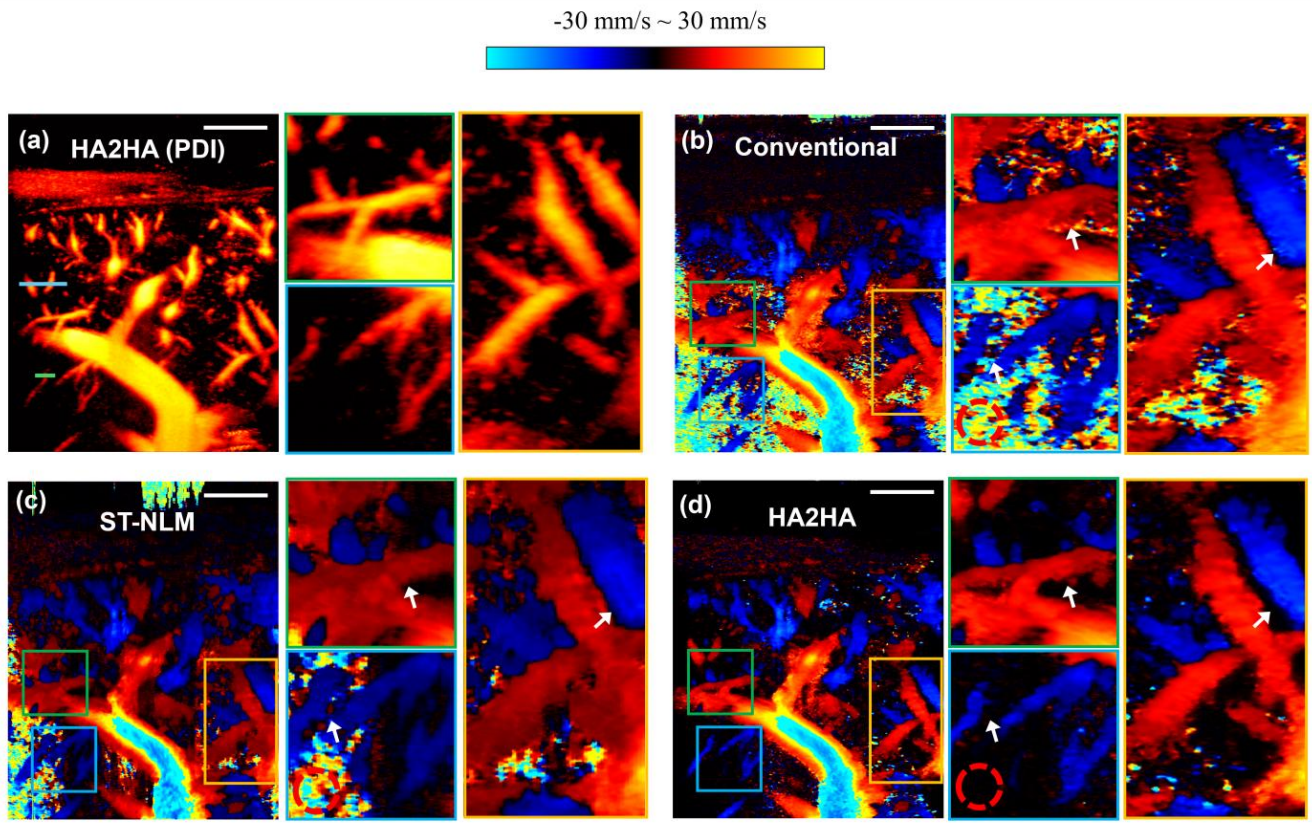

Fig. 9. (a) Power Doppler and (b-d) color Doppler imaging (CDI) results of *in-vivo* contrast-free human liver. (a) Power Doppler image obtained using the proposed HA2HA method, with green (vessel 1) and blue (vessel 2) lines indicating the cross-sectional locations of two vessels, whose intensity and velocity profiles are shown in Fig. 10. CDI results obtained using (b) the conventional method, (c) ST-NLM and (d) the proposed HA2HA method.  The white horizontal line represents 1 cm.

Despite its demonstrated effectiveness, the proposed method has several limitations and presents multiple directions for future exploration. Currently, it focuses exclusively on denoising. Incorporating super-resolution modules, such as those based on deconvolutional refinement, could further improve spatial resolution. Future work may also explore the use of structural priors to guide learning, as well as alternative data pairing strategies, for example by pairing different temporal frames or RF channels within a compounded acquisition. Furthermore, extending the framework to raw channel data denoising could enable more flexible system-level integration and facilitate broader deployment across various ultrasound systems.

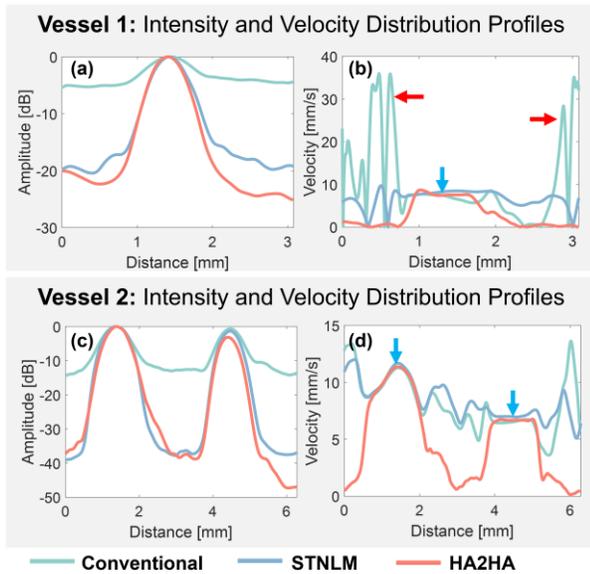

Fig. 10. Intensity and velocity distribution profiles corresponding to the two vessel locations marked in Fig. 9. (a) Intensity and (b) velocity profile of vessel 1 (green); (c) intensity and (d) velocity profile of vessel 2 (blue).

From a network design perspective, several directions can be explored to further improve denoising performance. First, extending the architecture from 2D to 3D convolutions could enable the model to better capture spatial-temporal dependencies across frames, especially in volumetric or time-resolved ultrasound acquisitions. Increasing network depth or adopting more advanced architectures, such as attention-based U-Nets [28] or transformer [40] backbones, may also enhance the representational capacity of the model.

While the current implementation is based on 2D RF data, the HA2HA framework can be naturally extended to 3D ultrasound imaging. The underlying strategy of constructing noise-variant yet signal-consistent training pairs remains applicable in volumetric acquisitions, offering potential for high-quality 3D microvascular imaging.

In addition, incorporating more sophisticated loss functions, such as adversarial loss, perceptual loss, or structure-aware penalties, may help preserve fine vascular structures while also enhancing the visual realism of the output. Another promising direction is to move beyond real-valued RF signals and operate directly in the complex domain using in-phase/quadrature (IQ) data. Leveraging IQ data may allow for reduced data size, thereby improving computational efficiency during both training and inference. Such improvements could make the method more scalable in real-time or resource-constrained applications.

## V. CONCLUSION

This study presents HA2HA, a self-supervised denoising framework for ultrasound microvascular imaging. By leveraging complementary angular subsets, HA2HA enables label-free training that suppresses noise while preserving vascular signals. By operating on RF signals, the method enables improved downstream processes such as color Doppler flow speed estimation based on RF phase information. Evaluated across diverse pig and human datasets, HA2HA demonstrates strong generalizability and consistently improves CNR, SNR, and CDI flow clarity, offering a scalable and clinically applicable solution for high-quality microvascular structural and functional imaging.